\documentclass[12pt,a4paper]{article}

\usepackage[utf8]{inputenc}
\usepackage[T1]{fontenc}
\usepackage{lmodern}
\usepackage[margin=2.5cm]{geometry}
\usepackage{setspace}
\usepackage{amsmath,amssymb}
\usepackage{graphicx}
\usepackage{booktabs}
\usepackage{natbib}
\usepackage{hyperref}
\usepackage{microtype}
\usepackage{caption}
\usepackage[british]{babel}
\hypersetup{
  colorlinks=true,
  linkcolor=black,
  citecolor=black,
  urlcolor=black
}

\title{Speed-Error Cross-Correlation Dating\\
       of Ancient Star Catalogues,\\
       with Application to the Almagest}

\author{Carlos \textsc{Baiget Orts}\thanks{Correspondence: \href{mailto:asinfreedom@gmail.com}{asinfreedom@gmail.com}. ORCID: \href{https://orcid.org/0009-0000-6725-5188}{0009-0000-6725-5188}. Family name: Baiget Orts.}\\
        \small\textit{Independent researcher, Valencia, Spain}}

\date{}

\begin{document}

\maketitle
\thispagestyle{empty}

\begin{abstract}
We present SESCC (Speed-Error Signals Cross-Correlation), a method for dating
ancient star catalogues from the cross-correlation between stellar proper-motion
speeds and positional residuals. At the true epoch, residuals are independent of
proper-motion speed; the epoch estimate is the trial date that minimises this
cross-correlation. For ecliptic latitudes, SESCC applies the dot product between
speeds and residuals across all catalogue stars without subset selection or linear
modelling. For ecliptic longitudes, SESCC-pairs uses pairwise longitude differences
between neighbouring stars, making the method immune to any global longitude offset
by algebraic construction. Validated against Tycho Brahe (1547\,CE, true
${\sim}$1580\,CE) and Ulugh~Beg (1452\,CE, true 1437\,CE), and confirmed invariant
under offsets of $\pm6^\circ$, the method is applied to the Almagest. Both
coordinates yield bootstrap distributions with 74\% pre-Christian minima, consistent
with a Hipparchan origin and inconsistent with a Ptolemaic one. The near-absence of
quarter-degree fractions in the Almagest longitudes, explained as the deterministic
consequence of Ptolemy's precession correction, provides independent corroboration.
\end{abstract}


\newpage

\section{Introduction}

The question of when an ancient star catalogue was observed has occupied
historians of astronomy for over three centuries. The ecliptic positions of
stars change over time due to proper motion, the slow individual displacement
of each star across the celestial sphere. A catalogue compiled at epoch~$T_0$
will show residuals --- differences between catalogued and modern positions
--- that are minimised when the trial epoch~$T$ matches~$T_0$. This is the
basis of all proper-motion dating methods.

For most historical star catalogues, the epoch of observation is known from
independent sources with sufficient precision for verification purposes. The
catalogue of Tycho Brahe reflects observations between 1576 and 1597\,CE
\citep{Dreyer1917}. The catalogue of Ulugh~Beg was compiled at his observatory
in Samarkand, with the epoch established as 1437\,CE from manuscript sources
\citep{Knobel1917, Verbunt2012}. These catalogues therefore serve as natural
validation benchmarks for any dating method: a method that fails to recover
their known epochs cannot be trusted when applied to catalogues of uncertain
date.

The Almagest star catalogue presents a different situation. Ptolemy assigns it
an epoch of 137\,CE, the reign of Antoninus Pius, but the catalogue's
systematic errors have led many investigators to question whether the positions
are genuinely Ptolemaic or derive instead from earlier observations by
Hipparchus of Nicaea, whose activity is documented between 162 and 127\,BCE.
The longitude errors show a systematic offset of approximately $-1^\circ$
relative to the Hipparchan epoch, consistent with Hipparchan positions adjusted to
137\,CE using Ptolemy's erroneous precession constant of $1^\circ$ per century
rather than the correct value of approximately $1.4^\circ$ per century.
This observation, noted since the time of Tycho Brahe and reviewed extensively by
\citet{Grasshoff1990}, motivates the application of proper-motion methods to
determine an independent epoch estimate for the catalogue coordinates.

The earliest systematic proper-motion dating of the Almagest was proposed by
\citet{Dambis2000}, whose `bulk method' applies linear regression between
position residuals and proper-motion speed to a selected group of fast-moving
stars, yielding $T = -89 \pm 112$ years. \citet{Duke2002} re-examined the bulk
method and demonstrated that the underlying linear correlation is not
statistically significant in the Almagest data. \citet{Nickiforov2009} further
showed that the method fails to recover the known epoch of Ulugh~Beg,
undermining its general validity. \citet{Brandt2014} used the declinations in
Book~VII of the Almagest rather than the catalogue proper, recovering an epoch
near $-128$\,BCE for observations attributed to Hipparchus. All these methods
address latitudes or declinations; longitude-based dating has relied on
precession arguments and offset analyses \citep{Dobler2002, Newton1977,
Newton1979} that necessarily incorporate assumptions about the global longitude
zero-point.

We present SESCC (Speed-Error Signals Cross-Correlation), a new proper-motion
dating method that addresses these limitations. At the true epoch, positional
residuals carry no systematic dependence on proper-motion speed; the epoch
estimate is the trial date that minimises this speed-error cross-correlation.
For ecliptic latitudes, SESCC applies the dot product between proper-motion
speeds and positional residuals across all catalogue stars simultaneously,
without selecting a subset and without fitting a linear model. For ecliptic
longitudes, we introduce an extension called SESCC-pairs that uses pairwise
longitude differences between neighbouring stars. Because any uniform longitude
offset --- whether from precession correction, instrumental zero-point error,
or any other global shift --- cancels exactly in the difference between any two
stars, SESCC-pairs is immune to global longitude offsets by construction. This
makes it the first longitude-based dating method that requires no assumption
about precession.

The paper is organised as follows. Section~\ref{sec:sescc} presents the SESCC
method for latitudes. Section~\ref{sec:pairs} introduces SESCC-pairs for
longitudes and establishes its invariance to global offsets.
Section~\ref{sec:validation} validates both methods against the catalogues of
Tycho Brahe and Ulugh~Beg, and against synthetic catalogues with realistic
positional noise. Section~\ref{sec:almagest} applies both methods to the
Almagest and discusses the results alongside the fractional position evidence.
Section~\ref{sec:conclusions} concludes.

\section{The SESCC Method for Ecliptic Latitudes}
\label{sec:sescc}

\subsection{Formulation}

Let a star catalogue contain $N$ entries, each identified with a star in the
Hipparcos catalogue \citep{ESA1997}. For each star~$i$, let $\beta_i$ be the
ecliptic latitude recorded in the catalogue, and let $\beta_i^{\rm mod}(T)$ be
the ecliptic latitude computed from the Hipparcos position and proper motion
propagated to trial epoch~$T$. The residual of star~$i$ at epoch~$T$ is
$\delta\beta_i(T) = \beta_i - \beta_i^{\rm mod}(T)$. The SESCC statistic is:
\begin{equation}
  C(T) = \sum_i |\mu_{\beta i}| \cdot |\delta\beta_i(T)|
  \label{eq:sescc}
\end{equation}
where $|\mu_{\beta i}|$ is the absolute latitudinal proper motion of star~$i$
in milli-degrees per millennium. The epoch estimate is
$\hat{T} = \arg\min_T C(T)$.

The statistical justification is as follows. At the true epoch~$T_0$, the
residual $\delta\beta_i(T_0)$ reflects the measurement error of star~$i$ in
the original catalogue, which is random and independent of $|\mu_{\beta i}|$.
The dot product $C(T_0)$ is therefore the sum of independent random terms with
no systematic tendency to grow with $|\mu_{\beta i}|$. At any other trial
epoch $T \neq T_0$, the residual contains an additional term proportional to
$|\mu_{\beta i}| \cdot |T - T_0|$, so fast-moving stars contribute
disproportionately large values to~$C(T)$. The statistic is therefore minimised
at the true epoch and increases as~$T$ departs from~$T_0$.

Stars with small $|\mu_{\beta i}|$ contribute little weight to the dot product
automatically, without any explicit selection or threshold. This implicit
weighting concentrates the signal in the stars that carry the most temporal
information, while stars with negligible proper motion contribute only noise
and are naturally down-weighted. Precession does not affect ecliptic latitudes,
so the method requires no precession correction.

\clearpage
\subsection{Practical implementation}

Modern positions and proper motions are taken from the Hipparcos catalogue
\citep{ESA1997}. Star positions at each trial epoch are computed using Skyfield
\citep{Rhodes2019} with the JPL DE441 ephemeris \citep{Park2021}.
Identifications between catalogue entries and Hipparcos stars follow
\citet{Verbunt2012}, whose machine-readable versions of the Almagest and
Ulugh~Beg catalogues provide ecliptic coordinates, positional errors relative
to Hipparcos, and identification quality flags. The trial epoch is scanned from 1900\,CE to $-600$\,BCE in steps of one year, covering approximately 2500 trial epochs. Positions at all epochs are
precomputed and stored in a compressed LZMA cache (\texttt{sescc\_positions.pkl.xz},
included in the code repository) to avoid redundant computation. If the cache
file is absent, positions are computed on the fly, which may require several
hours for a full 2500-year scan.

\subsection{Relation to previous approaches}

The bulk method of \citet{Dambis2000} selects a group of fast-moving stars and
fits a linear regression between their position residuals and proper-motion
speeds. SESCC differs in three fundamental respects. First, it operates on the
complete catalogue without any star selection, relying on implicit weighting by
$|\mu_{\beta i}|$ to concentrate signal in informative stars. Second, it
assumes no functional form for the relationship between residuals and
proper-motion speed --- only that the dot product is minimised at the true
epoch. Third, it requires no estimation of a systematic error component for
each star.

\citet{Duke2002} demonstrated that the linear correlation assumed by the bulk
method is not statistically significant in the Almagest data, and that the
regression is sensitive to which stars are included. Neither criticism applies
to SESCC: the dot product makes no linearity assumption, and the complete
catalogue is used. \citet{Nickiforov2009} showed that the bulk method fails to
recover the known epoch of Ulugh~Beg; Section~\ref{sec:validation} demonstrates
that SESCC recovers it correctly.

\subsection{Bootstrap resampling}

The epoch estimate~$\hat{T}$ from the full catalogue is a point estimate. To
characterise the sensitivity of the result to the composition of the stellar
sample, we apply bootstrap resampling: we draw $N$ stars with replacement from
the catalogue, compute the SESCC minimum for each resample, and repeat 1000
times. The resulting distribution characterises how much the result changes
when individual stars are added or removed.

We report the median, the 68\% range (16th to 84th percentile), and the
percentage of resamples giving a pre-Christian minimum. These are measures of
robustness, not classical inferential confidence intervals: they describe the
variability of the estimator under perturbations of the sample, not the
probability that the true epoch lies within a given interval.

\section{SESCC-pairs: Extension to Ecliptic Longitudes}
\label{sec:pairs}

\subsection{The problem of longitude dating}

Dating by ecliptic longitudes presents a difficulty that does not arise for
latitudes. Precession shifts all longitudes by the same amount --- approximately
$1.4^\circ$ per century --- so the modern position of a star depends on both its
proper motion and the accumulated precession since the epoch of observation. Any
longitude-based dating method must either correct for precession explicitly or
find a formulation that is immune to it.

The direct analogue of SESCC for longitudes --- applying the dot product between
longitudinal proper-motion speeds and absolute longitude residuals --- is
corrupted by this precession term. Furthermore, the Almagest longitudes carry
an additional global offset of approximately $-1^\circ$ whose origin is debated.
A method sensitive to the absolute value of longitude residuals will conflate
this offset with the temporal signal.

\subsection{Pairwise longitude differences}

SESCC-pairs resolves both issues by replacing absolute residuals with pairwise
differences. For each pair of stars $(i, j)$ satisfying proximity criteria (see
below), define the catalogue longitude difference $\delta\lambda_{ij} =
\lambda_i - \lambda_j$ and the modern longitude difference at trial epoch~$T$
as $\delta\lambda_{ij}^{\rm mod}(T) = \lambda_i^{\rm mod}(T) -
\lambda_j^{\rm mod}(T)$. The SESCC-pairs statistic is:
\begin{equation}
  C_{\rm p}(T) = \sum_{(i,j)} |\mu_{\lambda i}\cos\beta_i -
  \mu_{\lambda j}\cos\beta_j| \cdot
  |\delta\lambda_{ij} - \delta\lambda_{ij}^{\rm mod}(T)|
  \label{eq:pairs}
\end{equation}
where the sum runs over all pairs in the proximity set, and
$\mu_{\lambda i}\cos\beta_i$ is the angular speed of star~$i$ on the celestial
sphere in the direction of increasing longitude. The epoch estimate is
$\hat{T} = \arg\min_T C_{\rm p}(T)$.

The invariance to global longitude offsets follows directly from the pairwise
construction. If all longitudes are shifted by a constant $\varepsilon$, then
$\delta\lambda_{ij} = (\lambda_i + \varepsilon) - (\lambda_j + \varepsilon) =
\lambda_i - \lambda_j$, which is identical to the unshifted value. The same
cancellation applies to the modern positions. SESCC-pairs is therefore immune
to any global longitude offset by algebraic identity, not by approximation.

\subsection{Proximity criteria and pair selection}

Pairs are formed between stars satisfying $|\lambda_i - \lambda_j| <
\Delta\lambda_{\rm max}$ and $|\beta_i - \beta_j| < \Delta\beta_{\rm max}$.
We adopt default values of $\Delta\lambda_{\rm max} = 30^\circ$ and
$\Delta\beta_{\rm max} = 5^\circ$, which provide a balance between pair density
and proximity.

\subsection{Longitude positions in the J2000 inertial frame}

The modern longitude positions $\lambda_i^{\rm mod}(T)$ used in SESCC-pairs
are computed in the J2000 inertial ecliptic frame. This is essential: if
positions were computed in the ecliptic frame of date, the accumulated
precession would re-enter through the modern positions, corrupting the
cancellation that makes SESCC-pairs precession-independent.

\subsection{Stars with large parallax}

SESCC-pairs assumes that proper motion is linear over the 2500-year baseline
of the computation. We therefore exclude Keid (HIP~19849, $\pi = 198$\,mas)
and $\alpha$~Centauri (HIP~71681, $\pi = 742$\,mas), whose non-linear proper
motion introduces systematic biases that displace the epoch estimate by several
centuries. This exclusion applies only to SESCC-pairs; for SESCC applied to
latitudes the same stars are included without affecting the results.

\section{Validation}
\label{sec:validation}

\subsection{Validation strategy}

We validate SESCC and SESCC-pairs against three independent test cases: the
star catalogue of Tycho Brahe (known epoch $\sim$1580\,CE), the star catalogue
of Ulugh~Beg (known epoch 1437\,CE), and synthetic catalogues generated at
known epochs with realistic positional noise.

\subsection{Tycho Brahe}

The star catalogue of Tycho Brahe \citep{Verbunt2010} reflects observations
spanning approximately 1576 to 1597\,CE. SESCC applied to the full catalogue
(974 stars) gives $\hat{T} = 1570$\,CE for latitudes, 10 years from the
central observational epoch. SESCC-pairs applied to the 926 stars with
positional errors below 60 arcminutes gives $\hat{T} = 1547$\,CE for
longitudes (bootstrap median 1552\,CE, 68\% range [1535, 1582], std\,=
27\,yr). The longitude estimate lies within the documented observational span
of 1576--1597\,CE, and the bootstrap 68\% range brackets the entire span.

\clearpage
\subsection{Ulugh Beg}

The star catalogue of Ulugh~Beg \citep{Verbunt2012} has a precisely established
epoch of 1437\,CE. We apply both methods to the subset of 901 stars with
positional errors below 60 arcminutes. SESCC gives $\hat{T} = 1177$\,CE for
latitudes (bootstrap median 1177, 68\% range [1126, 1182], 0\% pre-CE).
SESCC-pairs gives $\hat{T} = 1452$\,CE for longitudes (bootstrap median 1454,
68\% range [1186, 1500], 0\% pre-CE), within 15 years of the true epoch.

The contrast between the latitude and longitude results for Ulugh~Beg is
informative. \citet{Verbunt2012} report that both coordinates show systematic
trends with ecliptic longitude --- the periodic patterns first identified by
\citet{Peters1915}. These spatially-variable systematic errors affect absolute
residuals (as used in SESCC) but cancel in pairwise differences between nearby
stars (as used in SESCC-pairs), explaining the superior performance of
SESCC-pairs on Ulugh~Beg.

\subsection{Synthetic catalogues}

To characterise the bias and precision of SESCC-pairs under controlled
conditions, we generate synthetic catalogues at known epochs with Gaussian
noise. The noise levels are taken directly from the empirical error
distributions of the Almagest as reported by \citet{Verbunt2012}: $\sigma
\approx 27'$ in longitude and $\sigma \approx 23'$ in latitude. We generate
20 realisations at each of two target epochs, $-127$\,BCE (Hipparchan) and
$+137$\,CE (Ptolemaic). Table~\ref{tab:synthetic} summarises the results.

\begin{table}[h]
  \centering
  \caption{SESCC-pairs validation results for synthetic catalogues (20 seeds each).}
  \label{tab:synthetic}
  \begin{tabular}{lcccc}
    \toprule
    Target epoch & Mean minimum & Std (yr) & 68\% range & \% pre-CE \\
    \midrule
    $-127$\,BCE & $-125$ & 179 & $[-304, -51]$ & 85\% \\
    $+137$\,CE  & $+148$ & 170 & $[-22, +307]$ & 20\% \\
    \bottomrule
  \end{tabular}
  \smallskip\\
  \footnotesize{Note: $\sigma = 23'$ (latitude), $\sigma = 27'$ (longitude),
  rounded to nearest $10'$. SESCC-pairs with $\Delta\lambda_{\rm max} = 30^\circ$,
  $\Delta\beta_{\rm max} = 5^\circ$, excluding HIP~19849 and HIP~71681.}
\end{table}

The mean bias is $+2$ years for the Hipparchan epoch and $+11$ years for the
Ptolemaic epoch. The two distributions are well separated: 85\% of Hipparchan
realisations give a pre-Christian minimum, compared to only 20\% of Ptolemaic
realisations. The invariance of SESCC-pairs to global longitude offsets is
verified empirically by applying the method to 13 versions of each catalogue
offset by integer degrees from $-6^\circ$ to $+6^\circ$; all 13 versions give
identical epoch estimates.

\section{Application to the Almagest}
\label{sec:almagest}

\subsection{Data}

We use the machine-readable version of the Almagest prepared by
\citet{Verbunt2012}, based on the edition of \citet{Toomer1984}. We retain the
1022 entries with secure or probable identifications and non-zero Hipparcos
numbers. The effective sample sizes are 1022 stars for SESCC and 844 stars for
SESCC-pairs after applying the positional error filter and excluding Keid and
$\alpha$~Centauri.

\subsection{Latitude dating}

SESCC applied to the full Almagest catalogue gives $\hat{T} = -49$\,BCE.
Bootstrap resampling (1000 resamples) gives a median of $-49$\,BCE, a 68\%
range of $[-177, +110]$, and 74\% of resamples with pre-Christian minima.
The dominant contributor is Arcturus (HIP~69673, $|\mu_\beta| = 629$
milli-deg/millennium), which accounts for a disproportionate fraction of the
total signal owing to its combination of high proper-motion speed and moderate
positional residual.
The 74\% pre-Christian fraction is the primary statistical result. Under the
null hypothesis that the catalogue is Ptolemaic, the synthetic catalogue
validation shows that 20\% of resamples give pre-Christian minima. The observed
74\% is therefore inconsistent with a Ptolemaic origin and consistent with a
Hipparchan origin (85\% in the synthetic validation). The point estimate of
$-49$\,BCE is compatible with the bulk-method result of $-89 \pm 112$ years
\citep{Dambis2000} and the declination-based estimate of $-128$\,BCE
\citep{Brandt2014}.

\begin{figure}[h]
  \centering
  \includegraphics[width=0.95\textwidth]{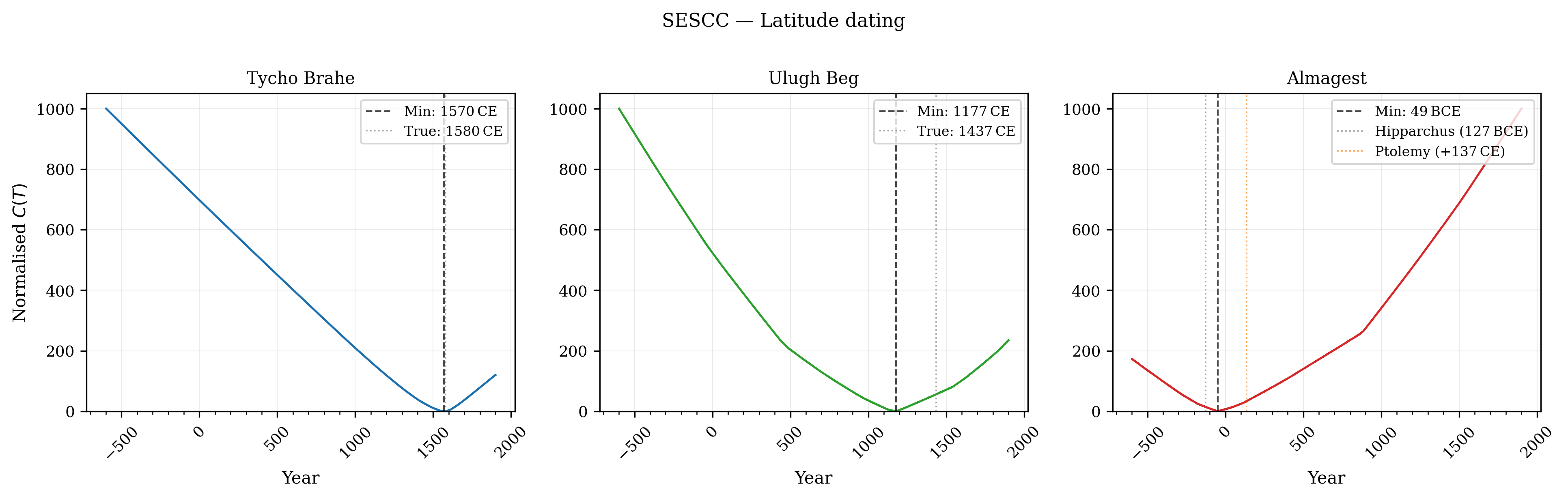}
  \caption{SESCC latitude dating curves $C(T)$, normalised to [0,1000], for
  Tycho Brahe (left), Ulugh~Beg (centre), and the Almagest (right). Vertical
  dashed lines indicate the known observational epoch (Brahe and Ulugh~Beg) or
  the Hipparchan ($-127$\,BCE) and Ptolemaic ($+137$\,CE) reference epochs
  (Almagest).}
  \label{fig:curves_lat}
\end{figure}

\subsection{Longitude dating}

SESCC-pairs applied to the filtered Almagest catalogue gives $\hat{T} =
-165$\,BCE for longitudes. Bootstrap resampling (1000 resamples) gives a median
of $-43$\,BCE, a 68\% range of $[-384, +176]$, and 74\% of resamples with
pre-Christian minima.

The agreement between the latitude result (74\% pre-CE) and the longitude
result (74\% pre-CE) is notable: the two methods use independent coordinates,
different stellar samples, and a fundamentally different computational strategy.
Their convergence on the same statistical conclusion strengthens the case for a
pre-Christian origin of both coordinate sets in the Almagest.

\begin{figure}[h]
  \centering
  \includegraphics[width=0.95\textwidth]{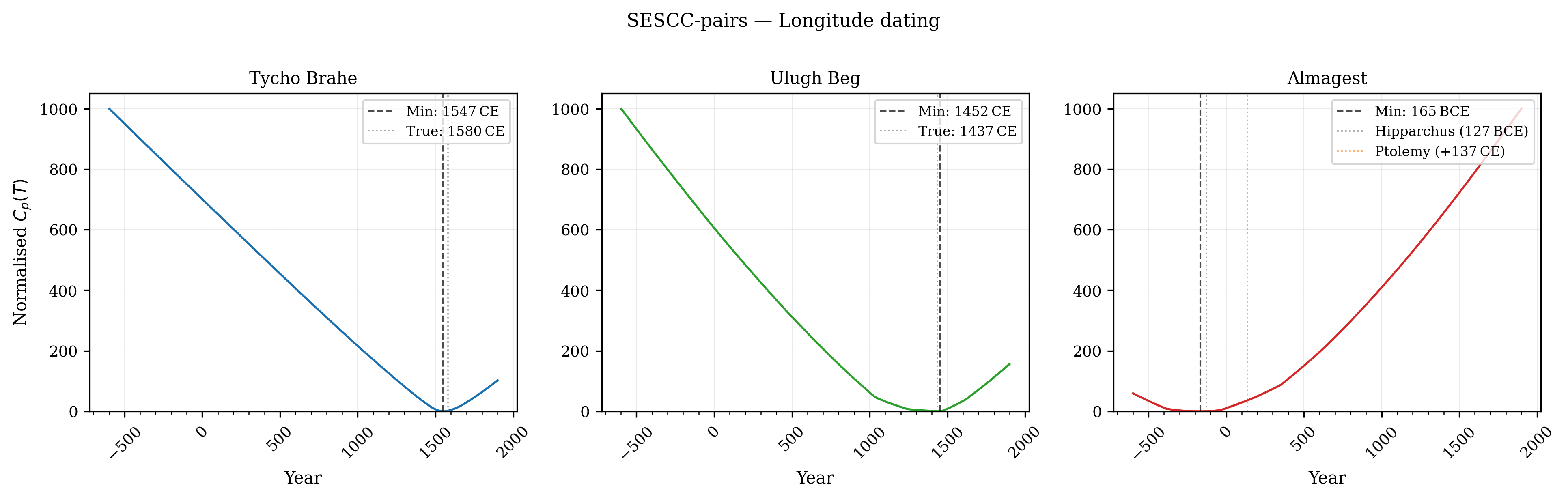}
  \caption{SESCC-pairs longitude dating curves $C_{\rm p}(T)$, normalised to
  [0,1000], for Tycho Brahe (left), Ulugh~Beg (centre), and the Almagest
  (right). Vertical dashed lines as in Figure~\ref{fig:curves_lat}.}
  \label{fig:curves_lon}
\end{figure}

\begin{figure}[h]
  \centering
  \includegraphics[width=0.95\textwidth]{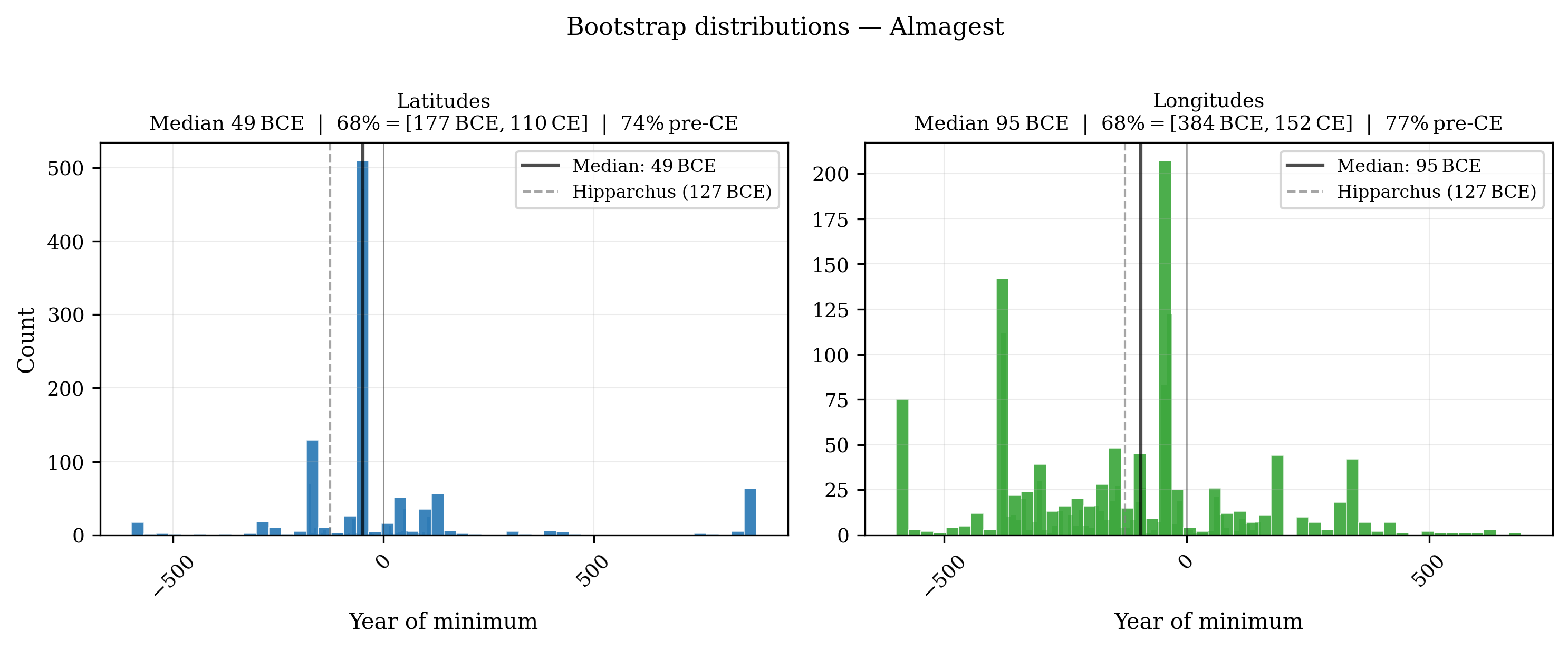}
  \caption{Bootstrap distributions of the epoch estimate for the Almagest,
  from SESCC latitudes (left) and SESCC-pairs longitudes (right). Vertical
  dashed lines mark the Hipparchan reference epoch ($-127$\,BCE) and the CE
  boundary.}
  \label{fig:bootstrap}
\end{figure}

\subsection{Evidence from sexagesimal fractions}
\label{sec:fractions}

The longitude dating result is consistent with the hypothesis that the Almagest
longitudes derive from Hipparchan observations adjusted by Ptolemy for
precession. If this is the case, one would expect to find structural evidence
of the arithmetic manipulation in the longitude data itself. Such a difference
exists and provides independent corroboration of the dating result.

The Almagest records positions in the sexagesimal system, with fractional parts
in multiples of $10'$ (system of sixths) or $15'$ (system of quarters).
\citet{Dreyer1917} noted that a subset of latitude entries use quarter-degree
fractions. \citet{Newton1979} showed that quarter-degree values ($M = 15', 45'$)
are nearly absent in the longitudes.

A coordinate-specific analysis reveals a striking asymmetry: quarter-diagnostic
fractions ($M = 15'$ or $45'$) account for 24.1\% of diagnostic latitude
entries but only 0.7\% of longitude entries ($\chi^2 = 173.3$, df\,= 1, $p \ll
0.001$; odds ratio\,= 44.2). This asymmetry is uniform across all magnitude
bins and all 48 Ptolemaic constellations, ruling out selective updating of any
particular stellar subset.

\begin{figure}[h]
  \centering
  \includegraphics[width=0.80\textwidth]{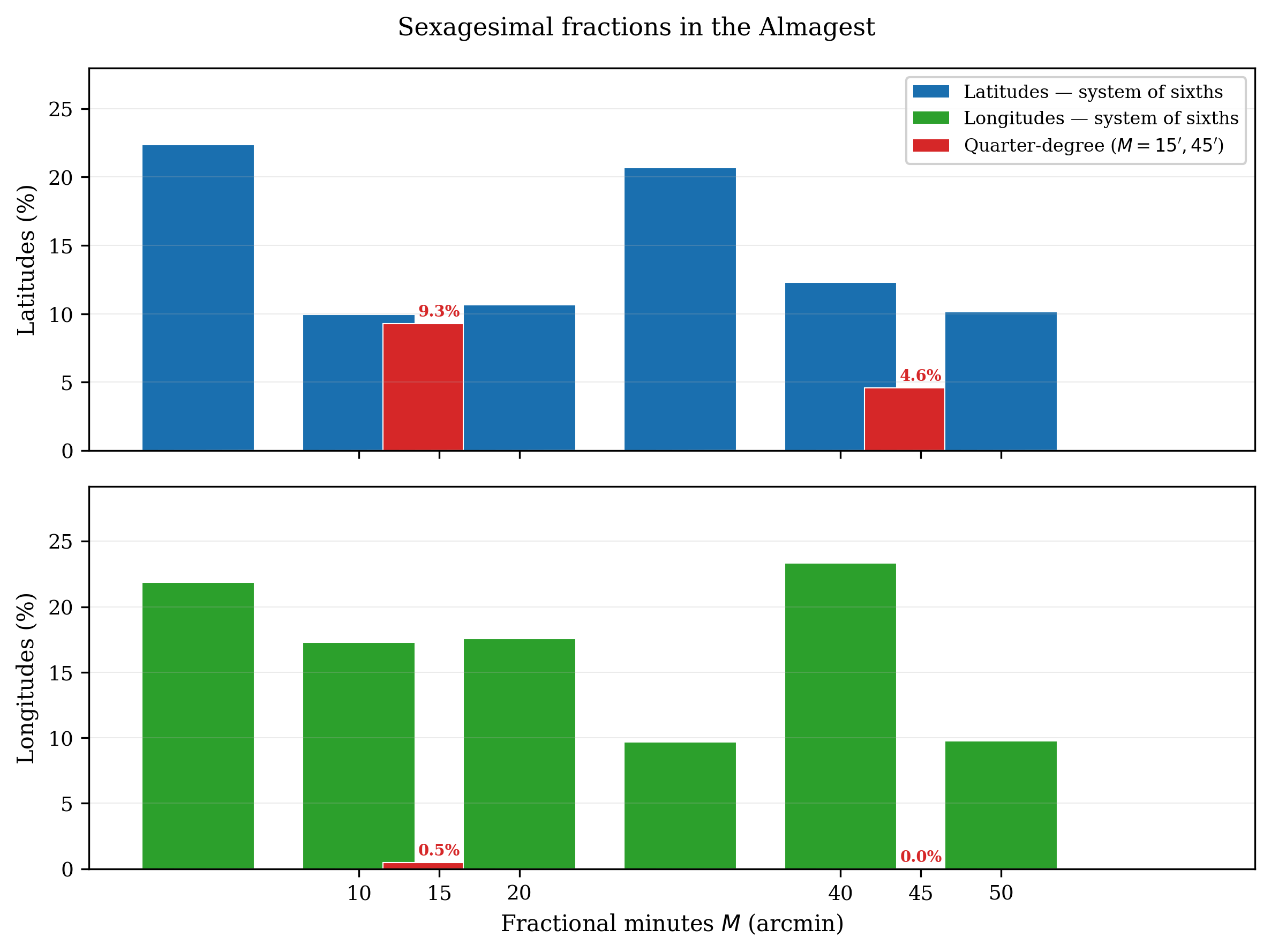}
  \caption{Distribution of sexagesimal fractional minutes in the Almagest for
  ecliptic latitudes (top) and ecliptic longitudes (bottom). The bars at
  $M = 15'$ and $M = 45'$ (quarter-degree fractions, shaded) are prominent in
  latitudes but nearly absent in longitudes.}
  \label{fig:fractions}
\end{figure}

The mechanism behind this asymmetry follows directly from Ptolemy's precession
correction. Ptolemy states explicitly in \emph{Almagest}~VII.3 that the
longitude difference between his epoch and that of Hipparchus amounts to
$2^\circ40'$, derived from his precession constant of $1^\circ$ per century over
265 years \citep{Toomer1984}. \citet{Newton1977} and \citet{Grasshoff1990}
discuss this correction in detail. Adding $2^\circ40'$ converts any fraction $M$
to $(M + 40') \bmod 60'$: $15' + 40' = 55'$ rounds to $50'$ or $60'$; $45' +
40' = 85' = 1^\circ25'$ rounds to $1^\circ20'$ or $1^\circ30'$. In neither case
does the result land on a quarter-diagnostic value. The addition of $2^\circ40'$
followed by rounding to the system of sixths therefore destroys the
quarter-fraction signature deterministically.

This provides a coherent account of the observed asymmetry. If Hipparchus
recorded both coordinates using an instrument with a quarter-degree graduation,
both would originally carry the quarter-fraction signature. Ptolemy copied the
latitudes unchanged --- he states that ``the latitudinal distances will remain
always unchanged'' (\emph{Almagest}~VII.3) --- and the quarter-fraction
signature survives to the present day. The longitudes were adjusted by adding
$2^\circ40'$, and the rounding erased the original fractional signature. This
provides an arithmetical explanation for the observation of \citet{Newton1979}
that quarter-degree fractions are nearly absent in the Almagest longitudes.

\section{Discussion and Conclusions}
\label{sec:conclusions}

We have presented SESCC, a proper-motion dating method based on the
cross-correlation between stellar proper-motion speeds and positional residuals.
The method operates on the complete catalogue without star selection, assumes
no functional form for the speed-residual relationship, and requires no
precession correction for ecliptic latitudes. Its extension SESCC-pairs uses
pairwise longitude differences to achieve exact algebraic immunity to global
longitude offsets, making it the first longitude-based dating method that
requires no assumption about precession.

Validation confirms the reliability of both methods. For Tycho Brahe, SESCC
gives 1570\,CE and SESCC-pairs gives 1547\,CE, both within or near the
documented observational span of 1576--1597\,CE. For Ulugh~Beg, SESCC-pairs
gives 1452\,CE, within 15 years of the true epoch. Synthetic catalogue
experiments show negligible bias and demonstrate that the Hipparchan and
Ptolemaic epochs are statistically distinguishable at the level of the Almagest
measurement noise.

Applied to the Almagest, both methods give consistent results. SESCC for
latitudes yields $\hat{T} = -49$\,BCE, with 74\% of bootstrap resamples
pre-Christian. SESCC-pairs for longitudes yields $\hat{T} = -165$\,BCE, also
with 74\% pre-Christian. The convergence of two independent methods on the same
statistical conclusion --- 74\% pre-Christian in both cases --- using different
coordinates, different stellar samples, and fundamentally different
computational strategies, constitutes the central result of this paper. Both
results are consistent with a Hipparchan origin and inconsistent with a
Ptolemaic origin.

The sexagesimal fraction evidence provides independent corroboration.
Quarter-degree fractions account for 24.1\% of diagnostic latitude entries but
only 0.7\% of longitude entries ($\chi^2 = 173.3$, odds ratio\,= 44.2). The
deterministic mechanism is the addition of $2^\circ40'$ followed by rounding:
any quarter-fraction longitude is converted to a non-quarter value, erasing the
original instrumental signature. The survival of quarter fractions in latitudes
and their absence in longitudes is precisely what the arithmetic of Ptolemy's
precession correction predicts if both coordinates were originally recorded by
Hipparchus.

Three limitations deserve emphasis. First, the precision of SESCC on the
Almagest is approximately $\pm200$ years (68\% bootstrap range), a consequence
of the catalogue's limited number of stars with sufficient proper motion.
Second, the bootstrap distribution for SESCC-pairs is wider than for SESCC,
reflecting the smaller effective sample size after filtering. Third, SESCC for
latitudes shows a systematic displacement of $-260$ years on Ulugh~Beg, likely
attributable to the spatially-variable systematic errors documented by
\citet{Peters1915}, which affect absolute residuals but cancel in pairwise
differences.

The source code implementing both methods is publicly available
\citep{SESCCcode}.

\section*{Acknowledgements}

This work depends fundamentally on open tools and open data. The author thanks
Brandon Rhodes for developing and maintaining Skyfield, which made precise
computation of historical stellar positions tractable. The statistical and
numerical analyses rely on NumPy \citep{Harris2020} and SciPy \citep{Virtanen2020}.
Frank Verbunt and Robert H.\ van Gent are gratefully acknowledged for publishing
machine-readable versions of the Almagest, Tycho Brahe, and Ulugh~Beg catalogues
with Hipparcos cross-identifications \citep{Verbunt2010, Verbunt2012}; without
these resources the SESCC analysis would not have been possible. This paper is
in part a testament to the impact of open science on historical research.

\noindent\textbf{Conflict of interest:} The author declares no conflict of
interest.

\noindent\textbf{Funding:} The author received no financial support for the
research, authorship, and/or publication of this article.

\bibliographystyle{plainnat}
\bibliography{references}

\end{document}